\documentclass[conference]{IEEEtran}

\IEEEoverridecommandlockouts
\usepackage{lineno,hyperref}
\usepackage{xcolor}
\usepackage{algorithm}
\usepackage{wrapfig}
\usepackage{booktabs}
\usepackage{multirow}
\usepackage{multicol}
\usepackage{subcaption}

\usepackage{cite}
\usepackage{amsmath,amssymb,amsfonts}
\usepackage{algorithmic}
\usepackage{graphicx}
\usepackage{textcomp}

\title{An Efficient Dynamic Multi-Sources To Single-Destination (DMS-SD) Algorithm In Smart City Navigation Using Adjacent Matrix
\thanks{This research was funded by: Key Program Special Fund in XJTLU under project KSF-E-64;
XJTLU Research Development Funding under projects RDF-19-01-14 and RDF-20-01-15; the National
Natural Science Foundation of China (NSFC) under grant 52175030.}
}

\author{\IEEEauthorblockN{Ziren Xiao}
\IEEEauthorblockA{\textit{Department of EEE} \\
\textit{Xi’an Jiaotong-Liverpool University}\\
Suzhou, China \\
ziren.xiao20@xjtlu.edu.cn}
\and
\IEEEauthorblockN{Ruxin Xiao}
\IEEEauthorblockA{\textit{Department of EEE} \\
\textit{Xi’an Jiaotong-Liverpool University}\\
Suzhou, China \\
ruxin.xiao21@student.xjtlu.edu.cn}
\and
\IEEEauthorblockN{Chang Liu}
\IEEEauthorblockA{\textit{Department of EEE} \\
\textit{Xi’an Jiaotong-Liverpool University}\\
Suzhou, China \\
chang.liu2020@outlook.com}
\and
\IEEEauthorblockN{Honghao Gao}
\IEEEauthorblockA{\textit{School of CES} \\
\textit{Shanghai University}\\
Shanghai, China \\
gaohonghao@shu.edu.cn}
\and
\IEEEauthorblockN{Xiaolong Xu}
\IEEEauthorblockA{\textit{Jiangsu Key Lab. of BBSIP} \\
\textit{NUPT}\\
Nanjing, China \\
xuxl@njupt.edu.cn}
\and
\IEEEauthorblockN{Shan Luo}
\IEEEauthorblockA{\textit{Department of CS} \\
\textit{University of Liverpool}\\
Liverpool, UK \\
shan.luo@liverpool.ac.uk}
\and
\IEEEauthorblockN{Xinheng Wang}
\IEEEauthorblockA{\textit{Department of EEE} \\
\textit{Xi’an Jiaotong-Liverpool University}\\
Suzhou, China \\
xinheng.wang@xjtlu.edu.cn}
}

\begin{document}

\maketitle

\begin{abstract}
Dijkstra's algorithm is one of the most popular classic path-planning algorithms, achieving optimal solutions across a wide range of challenging tasks. However, it only calculates the shortest distance from one vertex to another, which is hard to directly apply to the Dynamic Multi-Sources to Single-Destination (DMS-SD) problem. This paper proposes a modified Dijkstra algorithm to address the DMS-SD problem, where the destination can be dynamically changed. Our method deploys the concept of the Adjacent Matrix from Floyd's algorithm and achieves the goal with mathematical calculations. We formally show that all-pairs shortest distance information in Floyd's algorithm is not required in our algorithm. Extensive experiments verify the scalability and optimality of the proposed method.
\end{abstract}


\section{Introduction}
The emerging Industry 5.0 puts forward new requirements for the future intelligent transportation system. Based on the vision of Industry 4.0, beyond the autonomy, digitalization, self-awareness, connectivity, and interoperability in the future factory, Industry 5.0 also defines a new era of human-machine collaborative work that places sustainability, human care, and production resilience as the center\cite{xu2021industry}\cite{skobelev2017way}. As an essential part of a smart city, future smart transportation management, which coordinates, schedules, and monitors the city traffic, is expected to evolve following this wave in realizing the harmony among residents, urban, and society \cite{sharma2020industry}\cite{zichichi2020distributed}. To date, multiple enabling technologies, such as the internet of things, big data analytics, and artificial intelligence, drive the modern transportation network that converges and analyses transportation data to schedule and navigate \cite{jan2019designing}\cite{zichichi2020distributed}. In future Industry 5.0, smart transportation management must also reduce process waste and reinforce humanity's design while delivering adequate transportation. Thus, the future smart city is calling for new concepts, methods, and technologies to enhance its transportation management.  

Smart navigation system plays a decisive role in smart transportation management. As a coordinator in smart transportation management, smart navigation systems schedule and allocate residents' commutes to optimize the city traffic condition \cite{neil2020precise}. In Industry 5.0, smart navigation system tends to be increasingly important due to its influence on city productivity and residents' well-being. The vital point of the smart navigation system is the collaborative action analysis behind it. In real-time on-site situations, a smart navigation system must rapidly receive and respond to all users' states to derive an optimal traffic plan. Through the development of modern cities, the growing city population and vehicles have been introducing more commute units in transportation networks. Besides, the constructions of infrastructures bring in fluctuating road conditions. Therefore, complexity in transportation management keeps climbing rapidly. Thus, algorithms of collaborative action analysis must evolve abreast. 

Existing problems of collaborative action analysis remain unsolved in navigation research, and the Dynamic Multi Sources Single Destination problem (DMS-SD) is a critical one in transportation management. Congregate activities, such as sports competitions, outings, and business propagations, are arguably residents' most common daily affairs. Most congregate activities can only commence when all engagers have arrived. Therefore, while meeting at the same location, each engager's traveling time should be approximately the same to prohibit the waiting time of engagers who arrive earlier. Besides, the overall traveling time of each engager must be minimized to ensure the punctuality of congregate activities. As most road conditions in a city are generally the same, it can be hypothesized that traveling time is generally equivalent to traveling distance here. Moreover, on many occasions, congregate activities ought to temporarily change plans due to unpredicted issues (rain during outdoor activities, incomplete venue constructions, fire alarms, etc.). Thus, the original place must change to ensure congregate activities proceed, and so does the navigation destination.  

As a summary of such a problem in navigation terms, DMS-SD depicts a scenario in which multiple activity engagers need to unite at a specified location, where each engager travels equally and shortest distance. The intended destination may change dynamically depending on real-time on-site situations. Thus, route plans must be refreshed for every engager in correspondence. Meanwhile, as Industry 5.0 emphasizes process waste reduction, there are three additional requirements for DMS-SD solutions: (1) DMS-SD solutions should emphasize 'dynamic' so engagers can run the navigation system in real-time software with reasonable computation time; (2) Computation tasks of DMS-SD solutions should consume limited resources to run in engagers' mobile devices, rather than introducing enormous loads to a central server; (3) DMS-SD solutions should minimize the total distance to reduce power consumption generated from traveling. 

Conventionally, path planning algorithms simplify the problem into a point-to-point shortest time or shortest distance problem, defined as a Single Source to a Single Destination Problem (SS-SD). Based on this simplification, most existing approaches seek optimal routes according to performance indicators. For example, A* and Dijkstra algorithm can ensure an optimal solution via exploitation of entire path computations \cite{nosrati2012investigation}\cite{dijkstra1959note}. However, these methods are not effective enough to resolve DMS-SD problems as paths to all nodes must be simulated to obtain the shortest path in iterations, especially in a large-scale map. Besides, recent research also seeks solutions via Q-learning-based reinforcement learning algorithms \cite{abdi2021novel}\cite{liu2020mapper}. However, it is noticed to be challenging in DMS-SD problems due to the complexity of reward design. 

This paper proposes an innovative modified Dijkstra algorithm to address the DMS-SD problem. This modified algorithm introduces the Adjacent Matrix in classic Floyd's algorithm to store the shortest distance between nodes. Hence, this storage method avoids the mass computation work in Floyd's algorithm about distance calculations for every pair of nodes. On the base of adjustments brought by Adjacent Matrix, a temporary destination could only associate with the current locations of people in the DMS-SD problem. Therefore, this algorithm possesses the capacity for the optimal solution to the current state of people to navigate without a pre-training model. More importantly, individuals keep their personal preferences during decision-making. In this work, considering individual preference an influential factor, multiple preferences are provided to people, including distance, time, and meeting place, to satisfy different meeting goals rather than distance only. These beneficial changes enhance the completeness of problem-solving and fortify its user-friendly. The contributions of this paper lie in the following: 

\begin{itemize} 
    	\item Resolving the DMS-SD problem that has not been well discussed in past studies. 
        \item Proposing an efficient algorithm to address the DMS-SD problem based on Dijkstra's algorithm, which can effectively compute an optimal solution without generating the entire Adjacent Matrix. More importantly, the proposed method can address a large scaled DMS-SD problem with linearly increasing time. 
\end{itemize} 

The rest of the paper is organized as follows: Section 2 will clarify the details of the modified Dijkstra algorithm. Section 3 will present evaluations of the modified Dijkstra algorithm based on computation time, optimality, and dynamicity. Section 4 will review related work about multi-agent path planning to highlight the contribution and validation of this research. Section 5 will conclude the paper and discuss future relevant work.

\section{The Modified Dijkstra's (MD) Algorithm}\label{st:dijkstra}

Dijkstra's algorithm was proposed by \cite{dijkstra1959note} in 1959. This algorithm aims to search the shortest distance between a pair of vertices. In our work, we record all shortest distances from the source node to all other vertices using Adjacent Matrix, which is typically used in Floyd's algorithm. Then, by applying several matrix calculations, we achieve our goal: every user travels the shortest and equally distant to the destination based on the current state. The proposed algorithm spends much less time than Floyd's algorithm on solving the DMS-SD problem as well as efficiently generating an optimal solution.

\subsection{Adjacent Matrix}
The standard Adjacent Matrix ($M_{adj}$) is a 2-Dimensional matrix used to represent the connection between vertices in a graph. For example, provided a graph $G = (V,E)$ where the vertex set $V=\{v_1,v_2,v_3\}$ and $E$ is the edge set, the initial form of the $M_{adj}$ of $G$ can be presented as in Eq.~\ref{eq:example-madj}.

\begin{equation}
M_{adj} = 
\left[
{
\begin{array}{cccc}
 & v_1 & v_2 & v_3 \\ 
v_1 & 0 & 2 & 4 \\ 
v_2 & 2 & 0 & \infty \\ 
v_3 & 4 & \infty & 0 \\ 
\end{array}
} 
\right]
\label{eq:example-madj}
\end{equation}
where the notation $\infty$ means there is no direct road between those two vertices. $\infty$ can be updated to the actual number later if the destination can be achieved via another vertex. In our work, we use the Adjacent Matrix to present the shortest distance from a person to all other nodes. This means the matrix can be unsymmetric, as shown in Eq.~\ref{eq:example-madj-2}
\begin{equation}
M_{adj} = 
\left[
{
\begin{array}{ccccc}
 & v_1 & v_2 & v_3 & v_4 \\ 
A & 0 & 2 & 4 & 1\\ 
B & 2 & 0 & 6 & 3\\  
\end{array}
} 
\right]
\label{eq:example-madj-2}
\end{equation}
Where $A$ and $B$ are two people. Additionally, we use a tuple $(v_x,v_y)$ to represent the distance shown in the $M_{adj}$. For example, $(A,v_2)$ in Eq.~\ref{eq:example-madj-2} shows the distance from person $A$ to vertex $v_2$ is 2; $(A,v_1)$ indicates distance 1 because the person $A$ is at $v_1$.

\subsubsection{Initialise Adjacent Matrix}
The initial state of $M_{adj}$ only records the distance from people to their neighbours and themselves, while other distances are $\infty$. 

\subsection{Update Adjacent Matrix}
$M_{adj}$ is updated while exhausting the entire map. Firstly, we initialise a priority queue $Q$ sorted by the distance from the starting point (typically the person's current position) in ascending order. Initially, the $Q$ only contains the starting point $S$. We also maintain a list of visited vertices $L_v$, avoiding duplicate visitings. Then, the first vertex $n_i$ in the $Q$ is popped out, and each neighbour vertex of $n_i$ is examined. In specific, we update the distance from $S$ to the neighbour $n_{neighbour}$ if the new distance $S+n_i+n_{neighbour}$is smaller than $(S,n_{neighbour})$ in the $M_{adj}$. We repeat these operations until there is no vertex to be explored anymore in $Q$. The overview of how $M_{adj}$ is updated is summarised in the Algorithm.
 \ref{ag:dijkstra_algorithm}.

\begin{algorithm}
\caption{Dijkstra's Algorithm with Adjacent Matrix}
\begin{algorithmic}
\STATE initialise Adjacent Matrix $M_{adj}$
\STATE $N\leftarrow$ set of available nodes
\STATE $Q\leftarrow$ exploring priority queue
\STATE $L_{v}\leftarrow$ list of visited nodes
\STATE add the start node $S$ to $Q$
\WHILE{$Q$ is not $L_v$ empty}
\STATE $n_i\leftarrow$ pop an item from $Q$
\IF{$n_i$ is not in}
\STATE $L_{un}\leftarrow$ list of unvisited neighbours of $n_i$
\STATE update distance of $L_{un}$ from $Q$
\STATE add $L_{un}$ to $Q$
\STATE add $i$ to $L_{v}$
\ENDIF
\ENDWHILE
\end{algorithmic}
\label{ag:dijkstra_algorithm}
\end{algorithm}

\subsection{Extract Destination Node}
Since the updated $M_{adj}$ is optimal for the shortest distance between people and other vertices, we use the result of $M_{adj}$ to perform further calculations. Based on our targets, the extraction process can be two parts: (i) calculating the shortest total distance and (ii) every person travels a similar distance.

The first part $D_{total}$ is calculated by Eq.~\ref{eq:d-total}:
\begin{equation}\label{eq:d-total}
	D_{total} = \sum_{a\in people}M_{adj}^{a}
\end{equation}
where $M_{adj}^a$ is the row of Adjacent Matrix where the person $a$ is located. By summing up the from each person to a specific vertex, we can then choose the minimal distance from the $D_{total}$ row, which indicates the shortest total moving distance overall for all people. For example, the result of applying Eq.~\ref{eq:d-total} on the Adjacent Matrix in Eq.~\ref{eq:example-madj-2} is 
\begin{equation}
M_{adj} = 
\left[
{
\begin{array}{ccccc}
 & v_1 & v_2 & v_3 & v_4 \\ 
D_{total} & 2 & 2 & 10 & 4\\ 
\end{array}
} 
\right]
\label{eq:example-madj-2-total}
\end{equation}
It is very obvious that vertices $v_1$ and $v_2$ are with the shortest distance because person $A$ and $B$ are next to each other, in which either node costs a total distance of 2 to be achieved. 

For the second criterion, we use Eq.~\ref{eq:d-sim} to maintain a similar distance for every person:
\begin{equation}\label{eq:d-sim}
\begin{split}
D_{sim} = \sum_{a\in people}(\sum_{a'\in people}|M_{adj}^{a}-M_{adj}^{a'}|)\\
\end{split}
\end{equation}
We assume each node in $M_{adj}^a$ in Eq.~\ref{eq:d-sim} is a potential destination, thus, we calculate the distance difference from each node to all other people. For example, the result of applying Eq.~\ref{eq:d-sim} on the Adjacent Matrix in Eq.~\ref{eq:example-madj-2} is
\begin{equation}
M_{adj} = 
\left[
{
\begin{array}{ccccc}
 & v_1 & v_2 & v_3 & v_4 \\ 
D_{sim} & 2 & 2 & 2 & 2\\ 
\end{array}
} 
\right]
\label{eq:example-madj-2-sim}
\end{equation}

Then, all $D_x$s are normalised by Eq.~\ref{eq:norm} to avoid scaling problems, where the normalised value is between 0 and 1.
\begin{equation}\label{eq:norm}
	norm(x^i)=\frac{1-\frac{x^i}{\sum_{c\in x}c}}{2}
\end{equation}

Finally, $D_{total}$ and $D_{sim}$ are summed up with prioritised weights
\begin{equation}\label{eq:d}
    D = \alpha D_{total} + D_{sim}
\end{equation}
where $\alpha$ and $\beta$ show the corresponding weight of variables. For example, $D_{total}$ is more preferred (dominant), then $\alpha$ can be 0.9 and $\beta$ can be 0.1. The final solution is $argmin(D)$, extracting the node with minimal value as the goal destination.

\subsection{Multi-Objectives}\label{st:multiobjective}
The $M_{adj}$ currently represents the shortest distance, and we may use multiple $M_{adj}$ to demonstrate other metrics, e.g. time. The priority scores $S_{priority}^n$ are pre-defined by users, showing the weights of different preferences (0 to 5). Again, $S_{priority}^n$ are normalised by Eq.~\ref{eq:norm} to avoid scaling problems. For example, there are two objectives, distance and time, and two users set their preferences to $(4,5)$ on priority scores of distance $S_{priority}^{distance}$ and $(3,4)$ on priority scores of time $S_{priority}^{time}$; we assign the weight $0.56, 0.44$ to corresponding Adjacent Matrix: $M_{adj}^{distance}$ and $M_{adj}^{time}$ respectively. Eventually, the $M_{adj}$ is
\[M_{adj} = a M_{adj}^{distance} + b M_{adj}^{time}\]
where $a$ and $b$ are $0.56, 0.44$ in this example.

\subsection{Expected Performance}
The standard Dijkstra's algorithm needs $O(V+Elog(V))$ time to compute the result for each user, where $V$ is the number of nodes (vertices) and $E$ is the number of edges. The complexity can be enormous when there are many nodes, say 10,000. Consequently, it can cost time to compute the result, where users may not be satisfied with the response time. However, since Dijkstra's algorithm exhausts all possible solutions, the algorithm then guarantees an optimal solution, which is the critical advantage.

\begin{figure*}[!t]
\begin{multicols}{2}
 \includegraphics[width=0.51\textwidth]{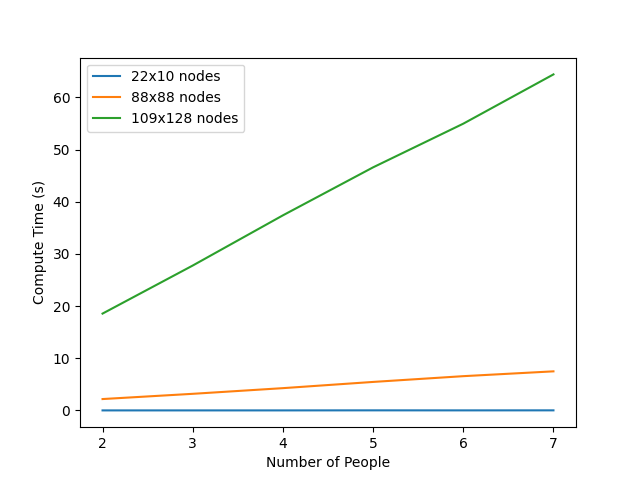}\par 
 \subcaption{The modified Dijkstra's algorithm's computation time}
 \label{fig:results_0a}
 \includegraphics[width=0.51\textwidth]{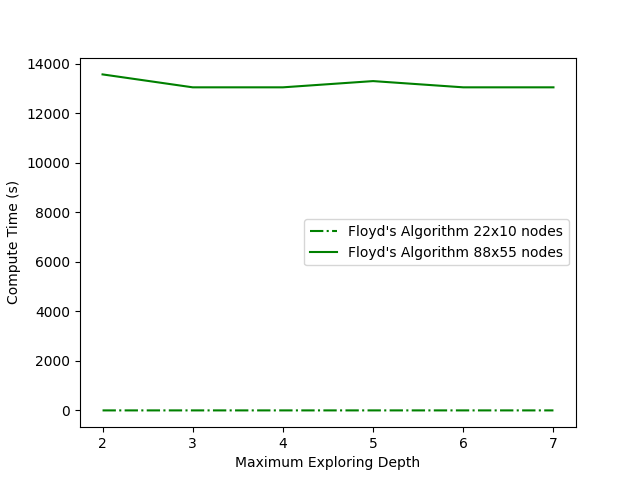}\par
 \subcaption{The Floyd's algorithm's computation time}
 \label{fig:results_0b}
\end{multicols}
\caption{The computation time with the change of the number of users in three size maps}
\label{fig:results_0}
\end{figure*}

\section{Evaluation}
\begin{figure*}
\begin{multicols}{5}
 \includegraphics[width=0.21\textwidth]{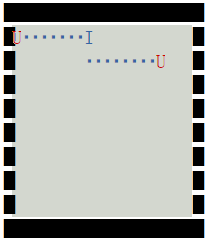}\par 
 \subcaption{2 users}
 \label{fig:results_1a}
 \includegraphics[width=0.21\textwidth]{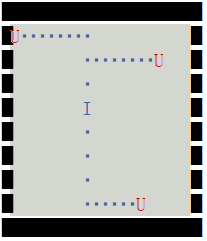}\par
 \subcaption{3 users}
 \label{fig:results_1b}
 \includegraphics[width=0.21\textwidth]{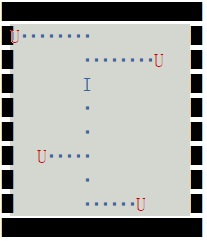}\par 
 \subcaption{4 users}
 \label{fig:results_1c}
 \includegraphics[width=0.21\textwidth]{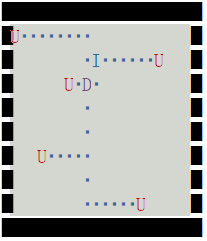}\par
 \subcaption{5 users}
 \label{fig:results_1d}
 \includegraphics[width=0.21\textwidth]{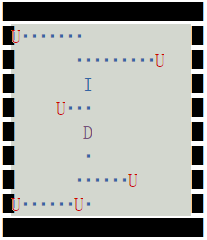}\par 
 \subcaption{6 users}
 \label{fig:results_1e}
\end{multicols}
\caption{Tests in a 22x10 map with different number of users}
\label{fig:results_1}
\end{figure*}

In this section, we evaluate the performance of the modified Dijkstra algorithm. We construct an environment that can load and render defined maps to test the algorithm. The map involves three elements, as shown in Fig.~\ref{fig:results_1}: (i) the wall (the black block), where users cannot pass through it; (ii) the accessible path (space); and (iii) the start point of each user (red letter U). Additionally, we use dots, D and I, representing the visited path, the achieved destination, and the initial destination, respectively. In the small map tests, there are only 22x10 vertices, and no walls are included inside the main map, as shown in Fig.~\ref{fig:results_1}. In another test, we simulate obstacles inside an 88x27 map to increase the difficulty of their movements. Both tests are terminated when all users meet at the same vertex. Our code\footnote{https://github.com/zirenxiao/maze-gym-rl/} for these simulations is open-sourced and implements our methods under the OpenAI Gym environment, which other researchers and practitioners can use to evaluate our approaches.

\begin{figure*}
\begin{multicols}{2}
\centering
 \includegraphics[scale=0.57]{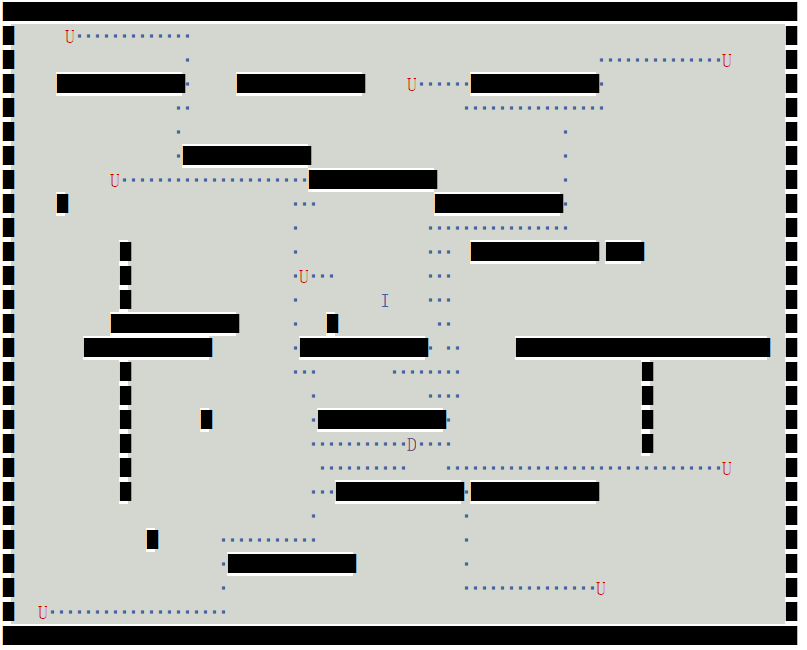}\par 
 \subcaption{8 users, random positions}
 \label{fig:results_2a}
 \includegraphics[scale=0.57]{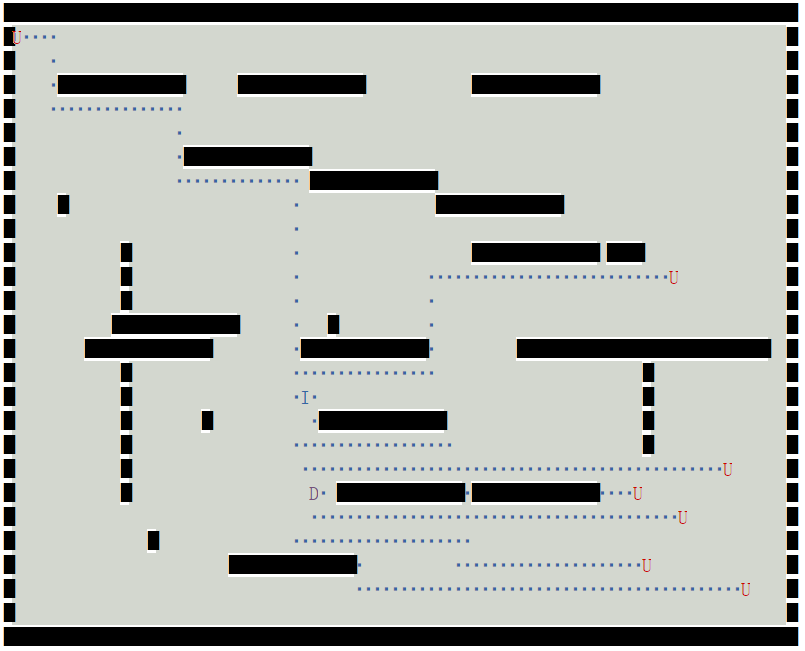}\par
 \subcaption{8 users, stick positions}
 \label{fig:results_2b}
\end{multicols}
\caption{Tests in a 88x27 map with different initial positions}
\label{fig:results_2}
\end{figure*}

\paragraph{Computation time} Fig.~\ref{fig:results_0a} shows the average computation time with the change in the number of users in three different size maps. In all three tests, it is evident that the computation time increases linearly when the number of users increases. Mainly, we observe that the computation time of the 109x128 map can be over the 60s when seven users are trying to find a common destination. This is because the modified Dijkstra's algorithm still needs to calculate the shortest distance from a user to all other vertices to find the optimal solution, which costs $O(V+Elog(V))$ time for each user. However, it only computes a partial Adjacent Matrix. More importantly, experiments show that our algorithm can be well scaled up to many vertices or users, where the time cost only increases linearly.
Floyd's algorithm relies heavily on the effectiveness of generating the entire Adjacent Matrix. Compared with Floyd's algorithm in Fig.~\ref{fig:results_0b}, our proposed method costs much less; the maximum computation time of Floyd's algorithm in our test can be several hours.

\paragraph{Optimality} We use the exhausting method to validate our first objective: if the total travel distance is minimized. The results of our tests show that our produced solutions are always optimal. 
However, there can be multiple solutions for a single state because of the objective, equal travel distance. For example, in Fig.~\ref{fig:results_1a}, the two user move 7 and 8 steps, respectively; we can let the left user moves eight steps, whereas another user moves 7 degrees, which still achieves both of our objectives.
We also observe that the algorithm tries to ensure that each user travels an equal distance. However, it is not fair for those users who are close to each other. For example, the right bottom users in Fig.~\ref{fig:results_2b} have to travel to the center of the map because the left top user is far away from those users.


\paragraph{Dynamicity} We note that the final destination is different from the initial calculation, such as in Fig.~\ref{fig:results_1d}, Fig.~\ref{fig:results_1e}, Fig.~\ref{fig:results_2a} and Fig.~\ref{fig:results_2b}, where $I$ is not overlapped with $D$. This is because there may exist multiple choices of movements for the current state. For example, a user can go up or down at a specific state, where both directions are the shortest distance to the position $argmin(D)$ calculated by Eq.~\ref{eq:d} and are towards the same destination. Although each movement for users is optimal for the global state, the destination for the current state can be changed and different from the initial one. This deviation can be magnified in a large map, such as in Fig.~\ref{fig:results_2a} and Fig.~\ref{fig:results_2b}. 
Consequently, in Fig.~\ref{fig:results_2a}, we can see wasted movements on the right side of I, which repeatedly moves to the left and right because the destination changes.

On the other hand, our algorithm can competently deal with state changes, including the shift in users' positions. It is beneficial when our algorithm is applied too frequently in changing environments, such as traffic jams, users' different moving speeds, and wrong path selection for a user.

\section{Related Work}
Research on multi-agent path planning for smart cities is still in its infancy. Search-based methods use mathematical formulas or strategies to calculate the final solution. Existing search-based studies can be roughly classified into two types: exhausting search and partial search. Early search-based methods, such as Dijkstra's algorithm~\cite{dijkstra1959note} and Floyd's algorithm~\cite{bellman1958routing}, visit every vertex in the search space and exhaust all possibilities to search for the best solution to an SS-SD problem. Although these can eventually generate an optimal solution, the computing time increases exponentially when the search space becomes bigger. A-STAR-Dijkstra-integrated algorithm~\cite{zhang2014multiple}, MPI~\cite{berger2015innovative} and Cooperative Conflict-Based Search (Co-CBS)~\cite{greshler2021cooperative} attempt to address MS-SD, where the destination is fixed and known. Compared with the original Dijkstra algorithm, these partial search-based approaches considerably reduce the computation time and promptly provide sub-optimal solutions for real problems because only heuristic-guided solutions are explored. Other variants of A* algorithm focusing on MS-SD with dynamically moving obstacles in the environment, such as Cooperative A* (CA*)~\cite{silver2005cooperative}, Hierarchical Cooperative A* (HCA*)~\cite{silver2005cooperative}, Windowed Hierarchical Cooperative A* (WHCA*) \cite{silver2005cooperative}, MAPP~\cite{wang2011mapp}, Multi-Robot D* Lite~\cite{al2011d} and Uncertainty M*~\cite{wagner2017path} algorithms implement heuristic functions to find the near-optimal path. Algorithms for searching collision-free trajectories in a multi-agent system were proposed in \cite{tahir2019heuristic} and \cite{li2019improved}. 
The existing search-based algorithms can solve SS-SD and potentially MS-SD planning problems based on fixed destinations. Due to partial area search, the heuristic-based search algorithms require less computational time to calculate the sub-optimal solution than entire-search algorithms. However, partial-search algorithms do not guarantee optimal solutions, causing inaccurate estimations of the destination; the error can be further magnified by multiple re-calculations of the destination and changes in users' state, which leads to a significant deviated destination. Additionally, partial-search methods require consideration of collision avoidance for motion planning.

\section{Conclusion}
In this paper, we propose a modified Dijkstra algorithm to address DMS-SD.
It achieves optimal solutions based on the mathematical calculations in Adjacent Matrix, especially fitting the requirements of DMS-SD, and can better meet user needs. Experiments conducted on the simulation environment demonstrate that the algorithm can effectively generate optimal solutions and achieve high scalability. However, in the extensive map test, we observe an unfairness in that the algorithm tries to ensure every user travels an equal distance, even for the isolated user and users who are very close. Therefore, as one of the future directions, we may consider that the isolated user can travel more distance to approach a group of users who are close enough instead of travelling a distance for all users equally.

\bibliographystyle{IEEEtran}
\bibliography{references}

\begin{thebibliography}{10}
\providecommand{\url}[1]{#1}
\csname url@samestyle\endcsname
\providecommand{\newblock}{\relax}
\providecommand{\bibinfo}[2]{#2}
\providecommand{\BIBentrySTDinterwordspacing}{\spaceskip=0pt\relax}
\providecommand{\BIBentryALTinterwordstretchfactor}{4}
\providecommand{\BIBentryALTinterwordspacing}{\spaceskip=\fontdimen2\font plus
\BIBentryALTinterwordstretchfactor\fontdimen3\font minus
  \fontdimen4\font\relax}
\providecommand{\BIBforeignlanguage}[2]{{%
\expandafter\ifx\csname l@#1\endcsname\relax
\typeout{** WARNING: IEEEtran.bst: No hyphenation pattern has been}%
\typeout{** loaded for the language `#1'. Using the pattern for}%
\typeout{** the default language instead.}%
\else
\language=\csname l@#1\endcsname
\fi
#2}}
\providecommand{\BIBdecl}{\relax}
\BIBdecl

\bibitem{xu2021industry}
X.~Xu, Y.~Lu, B.~Vogel-Heuser, and L.~Wang, ``Industry 4.0 and industry
  5.0—inception, conception and perception,'' \emph{Journal of Manufacturing
  Systems}, vol.~61, pp. 530--535, 2021.

\bibitem{skobelev2017way}
P.~Skobelev and S.~Y. Borovik, ``On the way from industry 4.0 to industry 5.0:
  From digital manufacturing to digital society,'' \emph{Industry 4.0}, vol.~2,
  no.~6, pp. 307--311, 2017.

\bibitem{sharma2020industry}
I.~Sharma, I.~Garg, and D.~Kiran, ``Industry 5.0 and smart cities: A futuristic
  approach,'' \emph{European Journal of Molecular \& Clinical Medicine},
  vol.~7, no.~08, pp. 2515--8260, 2020.

\bibitem{zichichi2020distributed}
M.~Zichichi, S.~Ferretti, and G.~D'Angelo, ``A distributed ledger based
  infrastructure for smart transportation system and social good,'' in
  \emph{2020 IEEE 17th Annual Consumer Communications \& Networking Conference
  (CCNC)}.\hskip 1em plus 0.5em minus 0.4em\relax IEEE, 2020, pp. 1--6.

\bibitem{jan2019designing}
B.~Jan, H.~Farman, M.~Khan, M.~Talha, and I.~U. Din, ``Designing a smart
  transportation system: An internet of things and big data approach,''
  \emph{IEEE Wireless Communications}, vol.~26, no.~4, pp. 73--79, 2019.

\bibitem{neil2020precise}
J.~Neil, L.~Cosart, and G.~Zampetti, ``Precise timing for vehicle navigation in
  the smart city: an overview,'' \emph{IEEE Communications Magazine}, vol.~58,
  no.~4, pp. 54--59, 2020.

\bibitem{nosrati2012investigation}
M.~Nosrati, R.~Karimi, and H.~A. Hasanvand, ``Investigation of the*(star)
  search algorithms: Characteristics, methods and approaches,'' \emph{World
  Applied Programming}, vol.~2, no.~4, pp. 251--256, 2012.

\bibitem{dijkstra1959note}
E.~W. Dijkstra \emph{et~al.}, ``A note on two problems in connexion with
  graphs,'' \emph{Numerische mathematik}, vol.~1, no.~1, pp. 269--271, 1959.

\bibitem{abdi2021novel}
A.~Abdi, D.~Adhikari, and J.~H. Park, ``A novel hybrid path planning method
  based on q-learning and neural network for robot arm,'' \emph{Applied
  Sciences}, vol.~11, no.~15, p. 6770, 2021.

\bibitem{liu2020mapper}
Z.~Liu, B.~Chen, H.~Zhou, G.~Koushik, M.~Hebert, and D.~Zhao, ``Mapper:
  Multi-agent path planning with evolutionary reinforcement learning in mixed
  dynamic environments,'' in \emph{2020 IEEE/RSJ International Conference on
  Intelligent Robots and Systems (IROS)}.\hskip 1em plus 0.5em minus
  0.4em\relax IEEE, 2020, pp. 11\,748--11\,754.

\bibitem{bellman1958routing}
R.~Bellman, ``On a routing problem,'' \emph{Quarterly of applied mathematics},
  vol.~16, no.~1, pp. 87--90, 1958.

\bibitem{zhang2014multiple}
Z.~Zhang and Z.~Zhao, ``A multiple mobile robots path planning algorithm based
  on a-star and dijkstra algorithm,'' \emph{International Journal of Smart
  Home}, vol.~8, no.~3, pp. 75--86, 2014.

\bibitem{berger2015innovative}
J.~Berger and N.~Lo, ``An innovative multi-agent search-and-rescue path
  planning approach,'' \emph{Computers \& Operations Research}, vol.~53, pp.
  24--31, 2015.

\bibitem{greshler2021cooperative}
N.~Greshler, O.~Gordon, O.~Salzman, and N.~Shimkin, ``Cooperative multi-agent
  path finding: Beyond path planning and collision avoidance,'' in \emph{2021
  International Symposium on Multi-Robot and Multi-Agent Systems (MRS)}.\hskip
  1em plus 0.5em minus 0.4em\relax IEEE, 2021, pp. 20--28.

\bibitem{silver2005cooperative}
D.~Silver, ``Cooperative pathfinding,'' in \emph{Proceedings of the aaai
  conference on artificial intelligence and interactive digital entertainment},
  vol.~1, no.~1, 2005, pp. 117--122.

\bibitem{wang2011mapp}
K.-H.~C. Wang and A.~Botea, ``Mapp: a scalable multi-agent path planning
  algorithm with tractability and completeness guarantees,'' \emph{Journal of
  Artificial Intelligence Research}, vol.~42, pp. 55--90, 2011.

\bibitem{al2011d}
K.~Al-Mutib, M.~AlSulaiman, M.~Emaduddin, H.~Ramdane, and E.~Mattar, ``D* lite
  based real-time multi-agent path planning in dynamic environments,'' in
  \emph{2011 third international conference on computational intelligence,
  modelling \& simulation}.\hskip 1em plus 0.5em minus 0.4em\relax IEEE, 2011,
  pp. 170--174.

\bibitem{wagner2017path}
G.~Wagner and H.~Choset, ``Path planning for multiple agents under
  uncertainty,'' in \emph{Twenty-Seventh International Conference on Automated
  Planning and Scheduling}, 2017.

\bibitem{tahir2019heuristic}
H.~Tahir, M.~N. Syed, and U.~Baroudi, ``Heuristic approach for real-time
  multi-agent trajectory planning under uncertainty,'' \emph{IEEE Access},
  vol.~8, pp. 3812--3826, 2019.

\bibitem{li2019improved}
J.~Li, A.~Felner, E.~Boyarski, H.~Ma, and S.~Koenig, ``Improved heuristics for
  multi-agent path finding with conflict-based search.'' in \emph{IJCAI}, vol.
  2019, 2019, pp. 442--449.

\end{thebibliography}

\end{document}